\documentclass[twocolumn]{aastex63}

\graphicspath{{V63/}}

\received{November 20, 2020}
\revised{January 22, 2021}
\accepted{February 4, 2021}
\submitjournal{ApJL}
\shorttitle{Hemispheric Tectonics on LHS 3844b}
\shortauthors{Meier et al.}
\usepackage{gensymb}
\usepackage{verbatim}
\usepackage{amsmath}
\usepackage{booktabs}

\begin{document}

\title{
Hemispheric Tectonics on super-Earth LHS 3844b
}
\correspondingauthor{Tobias G. Meier}
\email{tobias.meier@csh.unibe.ch}

\author[0000-0003-4143-8482]{Tobias G. Meier}
\affiliation{Center for Space and Habitability, University of Bern, Gesellschaftsstrasse 6, 3012 Bern, Switzerland}

\author[0000-0002-0673-4860]{Dan J. Bower}
\affiliation{Center for Space and Habitability, University of Bern, Gesellschaftsstrasse 6, 3012 Bern, Switzerland}

\author[0000-0002-3286-7683]{Tim Lichtenberg}
\affiliation{Atmospheric, Oceanic and Planetary Physics, Department of Physics, University of Oxford, Parks Road, Oxford OX1 3PU, United Kingdom}

\author[0000-0003-4878-621X]{Paul J. Tackley}
\affiliation{Institute of Geophysics, Department of Earth Sciences, ETH Zurich, Sonneggstrasse 5, Zurich 8092, Switzerland}

\author[0000-0002-9355-5165]{Brice-Olivier Demory}
\affiliation{Center for Space and Habitability, University of Bern, Gesellschaftsstrasse 6, 3012 Bern, Switzerland}

\begin{abstract}
The tectonic regime of rocky planets fundamentally influences their long-term evolution and cycling of volatiles between interior and atmosphere. Earth is the only known planet with active plate tectonics, but observations of exoplanets may deliver insights into the diversity of tectonic regimes beyond the solar system.  Observations of the thermal phase curve of super-Earth LHS 3844b reveal a solid surface and lack of a substantial atmosphere, with a temperature contrast between the substellar and antistellar point of around 1000 K.  Here, we use these constraints on the planet's surface to constrain the interior dynamics and tectonic regimes of LHS 3844b using numerical models of interior flow. We investigate the style of interior convection by assessing how upwellings and downwellings are organized and how tectonic regimes manifest. We discover three viable convective regimes with a mobile surface: (1) spatially uniform distribution of upwellings and downwellings, (2) prominent downwelling on the dayside and upwellings on the nightside, and (3) prominent downwelling on the nightside and upwellings on the dayside. Hemispheric tectonics is observed for regimes (2) and (3) as a direct consequence of the day-to-night temperature contrast. Such a tectonic mode is absent in the present-day solar system and has never been inferred from astrophysical observations of exoplanets. Our models offer distinct predictions for volcanism and outgassing linked to the tectonic regime, which may explain secondary features in phase curves and allow future observations to constrain the diversity of super-Earth interiors.
\end{abstract}

\section{Introduction} \label{sec:intro}

Plate tectonics is the unifying theory of Earth Science that explains the geological and surface evolution of Earth for at least the past 3 Gyr.  Plate tectonics is a fundamental component of long-term (geological) cycles that enable exchange of volatiles between the interior and atmosphere.  These cycles regulate climate and provide the necessary ingredients to nurture and sustain life on Earth and are thus essential to understanding the habitability of distant worlds.  The discovery of plate tectonics on Earth arose from seafloor mapping and seismology, but these techniques cannot be used to discern tectonic regimes on rocky exoplanets.  Instead, observations of the thermal phase curve are available for select planets.  Here, we link phase curve observations to numerical models of interior flow and constrain the possible tectonic regimes of super-Earth LHS 3844b.

Since the first thermal map of a super-Earth was constructed for super-Earth 55 Cnc e \citep{DGS12,DGW16}, other super-Earths have been targeted with observations to constrain their thermal emission and thus constrain their day- and nightside temperatures.  Recent efforts focused on LHS 3844b have found coincidence between the substellar point and the observed hotspot, and inferred a dayside temperature of $1040 \pm 40$ K and a nightside temperature around 0 K \citep{KKM19}.  These suggest that heat redistribution is inefficient; therefore the planet has neither substantial melt at its surface nor an active atmosphere.  The observational data suggests the emission originates from bare rock with a low albedo \citep{KKM19}.  Therefore, LHS 3844b has established itself as a case study for understanding the interior dynamics and tectonics of ultra-short period super-Earths.

Whether or not plate tectonics (mobile lid convection) operates on super-Earths has been a long-standing debate \citep[e.g.,][]{VOS06,OL07}, but the debate is notably hindered by the lack of observational constraints to supplement theoretical and numerical modeling efforts.  Tectonic regimes manifest from the interior dynamics of a planet, specifically the style and vigor of convection in the outermost silicate shell (the mantle).  Hence tectonic regimes are the surface expression of mantle flow that extends deep in the planet.  Therefore, introducing observational constraints into models is essential to provide new insights into the viable tectonic regimes operating on rocky exoplanets such as LHS 3844b.

It is unknown how the strong temperature contrast imposed by stellar irradiation on LHS 3844b influences its interior flow and hence its tectonic regimes.  Thus we utilize advanced models of interior flow coupled with thermal phase curve observations to determine the viable convective regimes operating in the interior of an ultra-short period super-Earth.  This enables us to probe the coupling between the surface and interior of LHS 3844b and thereby infer observational strategies for further geological characterization.
\section{Constraining interior dynamics with observations} \label{sec:model}
LHS 3844b is 1.3 Earth radii \citep{VHV18} and the dayside and nightside temperatures are 1040 K and $\approx 0$ K, respectively \citep{KKM19}.  We estimate the longitudinal temperature variation by assuming a blackbody at equilibrium and that only the dayside reradiates energy received from the star. Since the thermal phase curve only constrains the average longitudinal dependence of surface temperature, we construct models of interior flow within 2D spherical annulus geometry \citep{HTACK08}.  Mantles behave as highly viscous fluids, so we solve for Stokes flow (mass, momentum, and energy conservation) using the mantle convection code StagYY \citep{TACK08}.  Since the mass of LHS 3844b has not been measured, we assume its mantle is silicate rock and has the same relative thickness as Earth.  We adopt a hydrostatic reference state to provide the mantle profiles of material properties \citep{TAB13}.  Utilizing an Arrhenius-type viscosity law, the mantle is modeled with an upper mantle, a perovskite layer, and a post-perovskite layer assuming a lower bound estimate of the viscosity \citep{TAB13}.  The perovskite-post-perovskite interface occurs at a depth around $1680$ km (total mantle depth is $\approx$3500 km).

Lithospheric strength is modeled by a plastic yielding criteria ($\sigma_{\rm duct}$) to obtain self-consistent plate-like behavior \citep{MS98, T00}. Laboratory experiments estimate the yield stress to be a few hundred MPa \citep{KEM95} and numerical simulations employ a yield stress less than $\sim 150$ MPa to obtain plate-like behavior for Earth-like planets \citep{T00}.  We vary the ductile yield stress between 10 MPa and 300 MPa to model a lithosphere that is weak and strong, respectively.  Planetary mantles can be heated from below from cooling of an iron-rich (geophysical) core and heated from within from decay of radionuclides or tidal heating.  We therefore explore two heating modes: (1) basal heating, where the mantle is heated exclusively by the geophysical core, and (2) mixed-mode heating, where constant mantle heat production (Earth-like at $5.2\times10^{-12}$ W kg$^{-1)}$ supplements heating by the core. \cite{KRU20} estimate the age of the host star as $7.8 \pm 1.6$ Gyr, but an estimate of the stellar radionuclide abundances is presently not available to provide a constraint on the radiogenic heat budget of the planet \citep[e.g.,][]{UJP15}.  Therefore, we assume an Earth-like internal heat budget, in accordance with previous modeling efforts \citep[e.g.,][]{KRU20}.

\section{Tectonic regimes and interior flow}
\begin{figure*}
    \centering
    \includegraphics[width=0.99\textwidth]{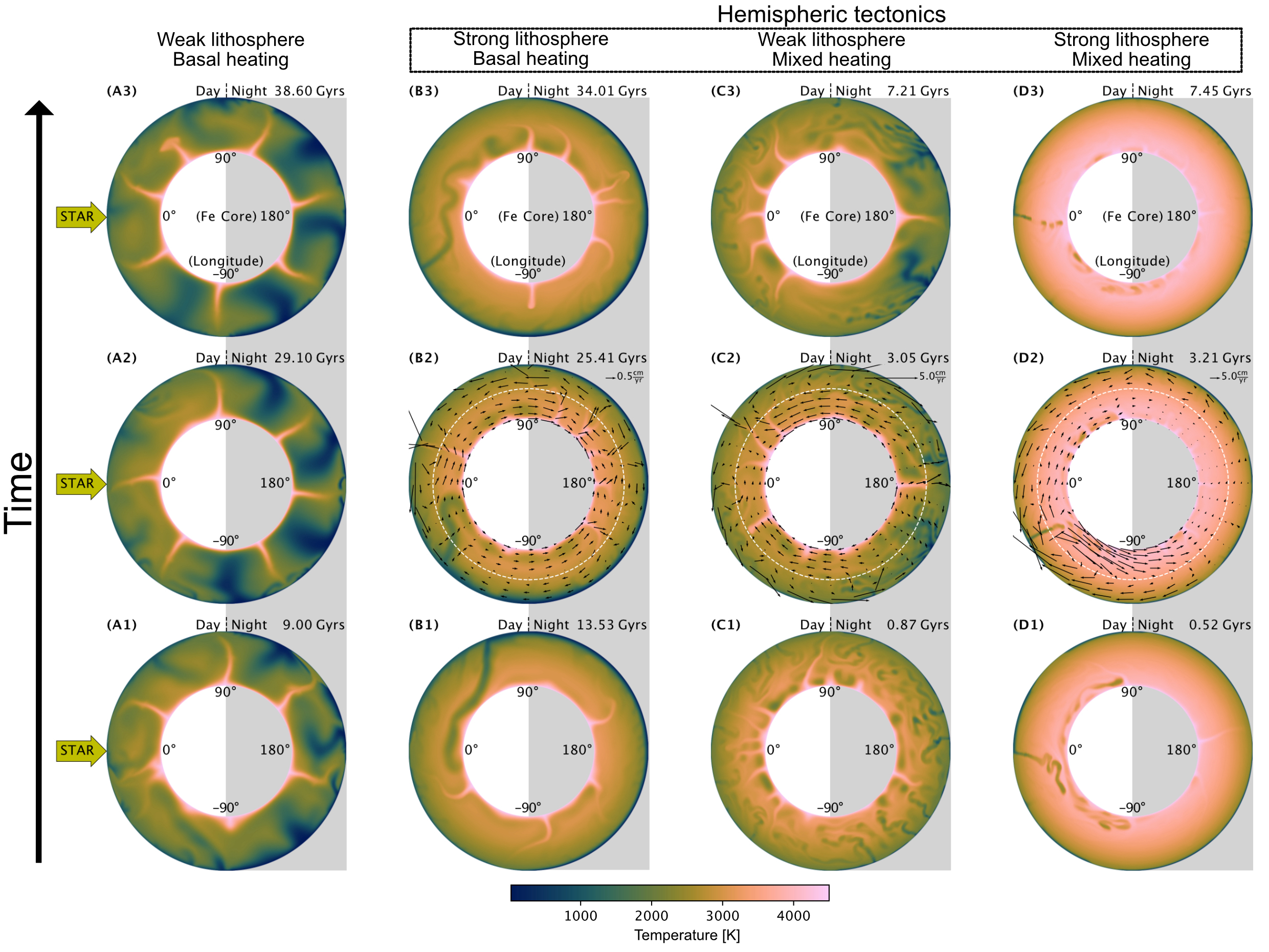}
    \caption{Mantle temperature for tectonic regimes of LHS 3844b for (A) weak lithosphere ($\sigma_{\rm duct}$= 10 MPa) and basal heating, (B) strong lithosphere ($\sigma_{\rm duct}$= 300 MPa) and basal heating, (C) weak lithosphere ($\sigma_{\rm duct}$= 10 MPa) and mixed heating, and (D) strong lithosphere ($\sigma_{\rm duct}$= 300 MPa) and mixed heating.  Upper left label in each cross-section (A3, B3, etc.) indicates the time of the cross-section in Figure~\ref{fig:zebra_compact}. The substellar point is at 0$^\circ$ longitude and the nightside (90$^\circ$ -- 270$^\circ$) is denoted by a gray background.  White dashed line in the mid-mantle in B2, C2, and D2 shows the boundary between perovskite (low pressure) and post-perovskite (high pressure).}
    \label{fig:evo_compact}
\end{figure*}

Our simulations discover three mobile lid tectonic regimes, each of which is associated with a distinct interior temperature and flow field: (1) uniform distribution of upwellings and downwellings (Figs.~\ref{fig:evo_compact}A and \ref{fig:zebra_compact}A), (2) downwellings on the dayside and upwellings on the nightside (Figs.~\ref{fig:evo_compact}B, D and \ref{fig:zebra_compact}B, D), and (3) downwellings on the nightside and upwellings on the dayside (Figs.~\ref{fig:evo_compact}C and \ref{fig:zebra_compact}C).  Figure~\ref{fig:evo_compact} shows the mantle temperature for three times of each model where the substellar point is located at 0 degrees. Figure~\ref{fig:zebra_compact} shows evolutionary tracks that summarize the distribution of upwellings (red tracks) and downwellings (blue tracks) with longitude and time. The time of the corresponding temperature fields in Fig.~\ref{fig:evo_compact} is indicated by horizontal dashed lines in Fig.~\ref{fig:zebra_compact}.  The internal heating ratio \citep[e.g.,][]{K17} is zero (by definition) for the basally heated models and between 0.7 and 0.8 for the internally heated cases. The time-averaged core--mantle boundary (CMB)  heat flux is $\approx 70$ mW m$^{-2}$ and $\approx 50$ mW m$^{-2}$ for the internally heated weak and strong lithosphere models, respectively. For the basally heated models, the CMB heat flux is $\approx 95$ mW m$^{-2}$ and $\approx 65$ mW m$^{-2}$ for the weak and strong lithosphere models, respectively.

\subsection{Uniform distribution of upwellings and downwellings}
A uniform distribution of upwellings and downwellings is predicted if the planet is dominantly basal heated and has a weak lithosphere (Figs.~\ref{fig:evo_compact}A and \ref{fig:zebra_compact}A). However, downwellings are stronger on the nightside than the dayside because the upper thermal boundary layer imposes a larger temperature contrast between the surface and the interior (Fig.~\ref{fig:evo_compact}A). This results in strong downwellings since the contrast determines both thermal buoyancy as well as material strength through viscosity. High viscosity downwellings dictate the long-wavelength pattern of flow and thus shepherd the upwellings (also known as plumes, orange/pink structures in Fig.~\ref{fig:evo_compact}A) along the core--mantle boundary (CMB) into position between the downwellings.  The pinning of plumes by downwellings is evident by the longitudinal stability of the upwelling and downwelling tracks over several Gyrs (Fig.~\ref{fig:zebra_compact}A).

Both the downwellings and the upwellings exhibit little lateral migration in longitude, only oscillating back and forth by 45--60 degrees.  Low viscosity upwellings mirror the migration of high viscosity downwellings, maintaining a constant separation at all times (Fig.~\ref{fig:zebra_compact}A).  This occurs for the dayside and the nightside, even though the downwellings are weaker on the dayside.  The evolutionary tracks in Fig.~\ref{fig:zebra_compact}A also reveal that upwellings and downwellings are maintained by a constant draining of the upper and lower thermal boundary layer, respectively.  Hence once the arrangement of downwellings and upwellings is established in the interior, they tend to persist and no new boundary layer instabilities occur.  Therefore both the day- and nightside only exhibit a small amount of time-dependent flow.
\subsection{Hemispheric tectonics: downwellings on dayside}

\begin{figure*}
    \centering
    \includegraphics[width=0.99\textwidth]{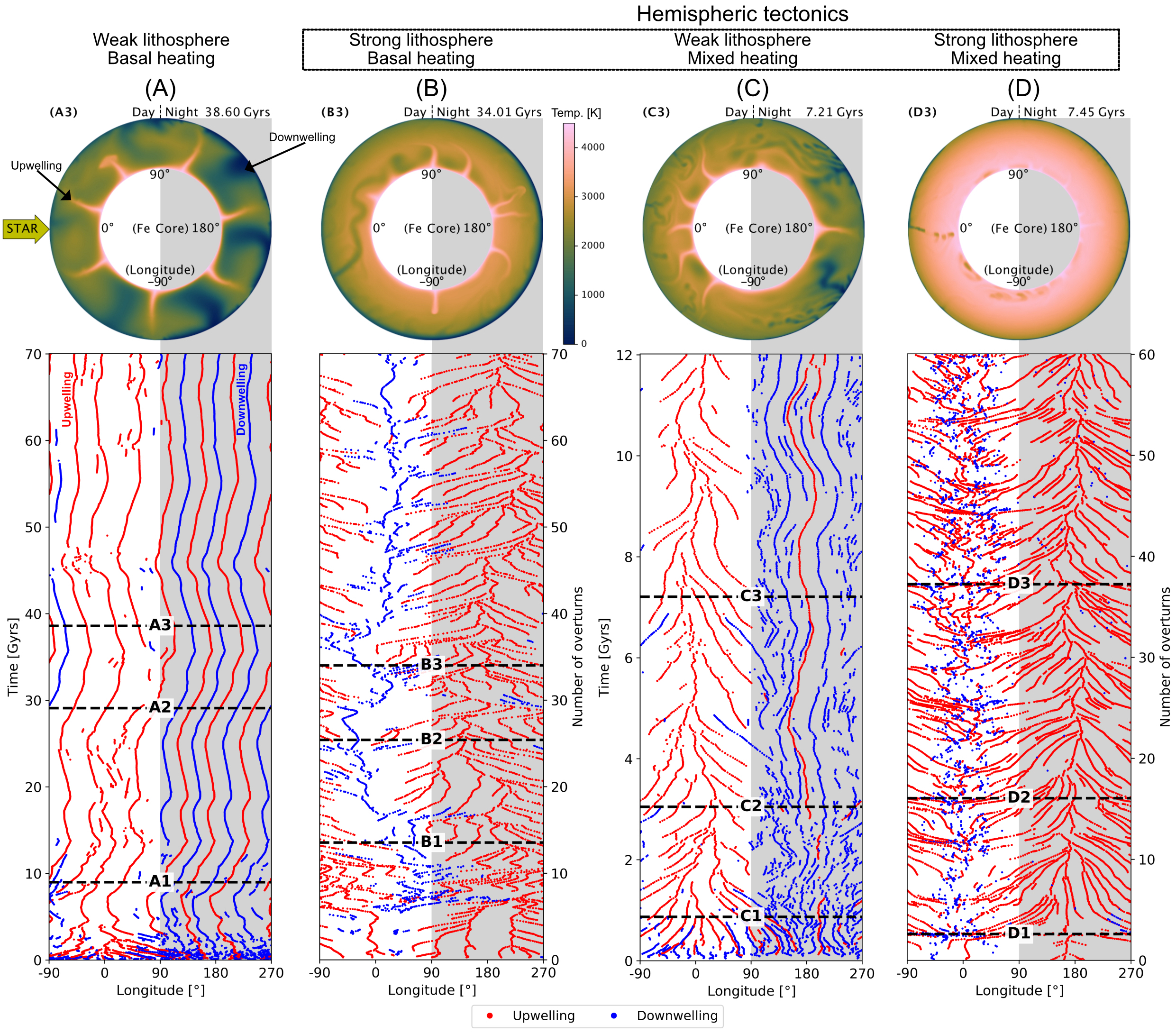}
    \caption{Evolutionary tracks for tectonic regimes of LHS 3844b.  Cross-sections (A3, B3, C3, and D3) show arrangement of upwellings and downwellings in the mantle.  Below, the evolution charts show the longitude of upwellings (red) and downwellings (blue) as a function of time.  Horizontal dashed lines denote the times of the cross-sections including those in Fig.~\ref{fig:evo_compact}. The substellar point is at 0$^\circ$ longitude and the nightside (90$^\circ$ -- 270$^\circ$) is denoted by a gray background.}
    \label{fig:zebra_compact}
\end{figure*}

\begin{figure*}
    \centering
    \includegraphics[width=0.99\textwidth]{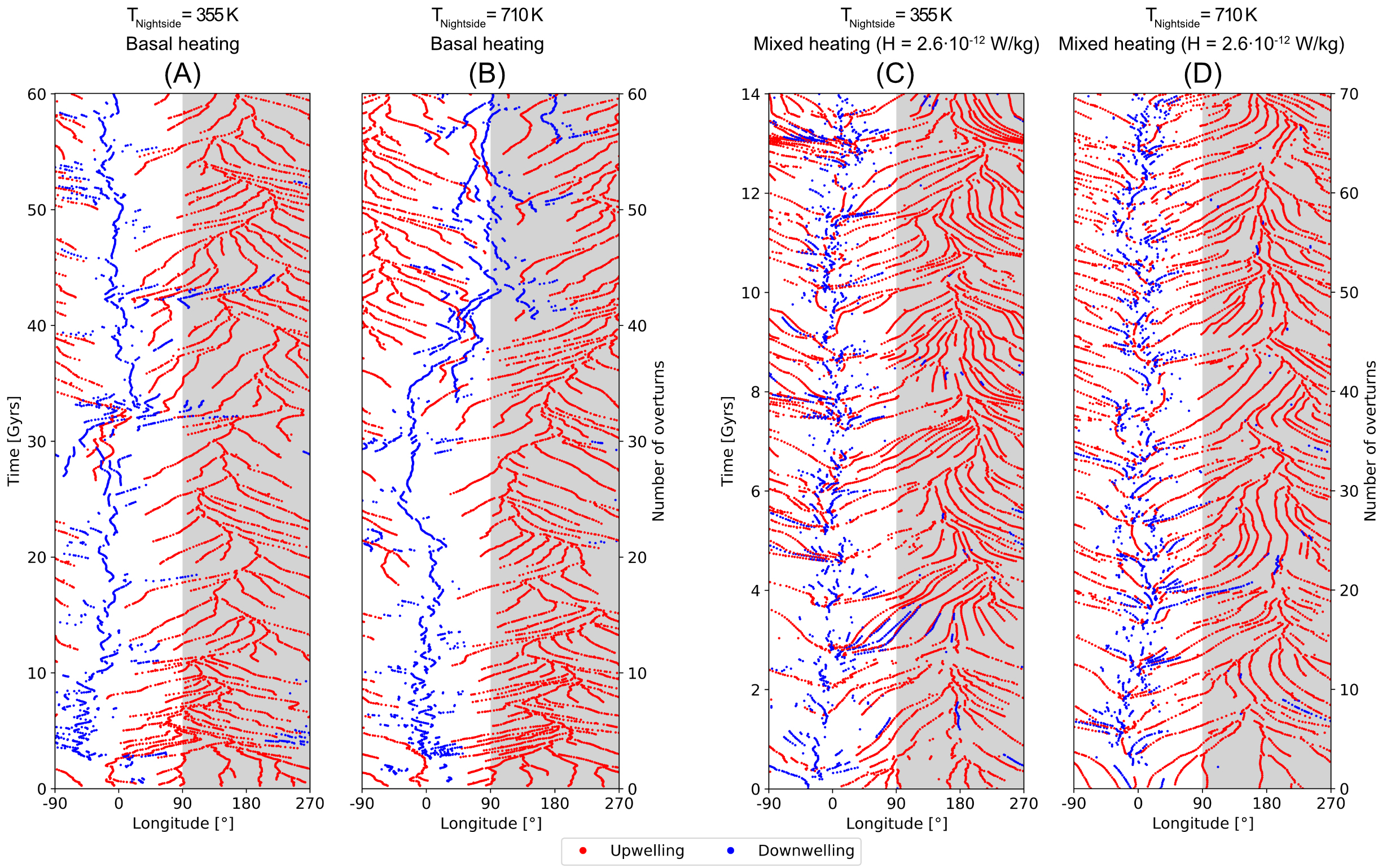}
    \caption{Evolutionary tracks for the strong lithosphere case with a higher nightside temperature (based on observational uncertainty) and reduced internal heating rate compared to the reference cases (Fig.~\ref{fig:zebra_compact}: B, D).  (A) basal heating and $T_{\rm Nightside}=355$ K, (B) basal heating and $T_{\rm Nightside}=710$ K, (C) mixed heating (half the reference rate) and $T_{\rm Nightside}=355$ K, and (D) mixed heating (half the reference rate) and $T_{\rm Nightside}=710$ K.  Hemispheric tectonics persist throughout this parameter variation.}
    \label{fig:zebra_compact_newmodels}
\end{figure*}

A dominant downwelling on the dayside and upwellings on the nightside are predicted with a strong lithosphere, independent of the heating mode (Figs.~\ref{fig:evo_compact}B, D and \ref{fig:zebra_compact}B, D).  A degree-1 convection pattern is established and hence hemispheric tectonics operates at the surface.  Figures~\ref{fig:evo_compact}B and \ref{fig:zebra_compact}B show the temperature for the basally heated model with a strong lithosphere.  A prominent downwelling forms on the dayside and descends into the deep mantle (Fig.~\ref{fig:evo_compact}: B1). It flushes hot material from the lower thermal boundary layer around the CMB from the dayside to the nightside, thereby promoting plumes on the nightside that rise to the surface (Fig.~\ref{fig:evo_compact}: B2, B3).  This deep mantle flow from dayside to nightside is accommodated in the post-perovskite layer, whereas the return flow is established in the perovskite layer above and delivers cold material from the nightside to the dayside (Fig.~\ref{fig:evo_compact}: B2).  The near-surface advection of cold material from the nightside to the dayside thickens the upper thermal boundary layer on the dayside, which helps to sustain the prominent downwelling.

Figure~\ref{fig:zebra_compact}B shows that the downwelling (blue track) remains close to the substellar point (0$^\circ$) rather than to the day--night terminator (90$^\circ$ or -90$^\circ$).  The downwelling displaces hot material laterally along the CMB, thereby promoting the thickening of the lower thermal boundary layer and hence plume formation.  This leads to the initiation of upwellings on the dayside that are pushed toward the nightside (Fig.~\ref{fig:evo_compact}: B3).  This is evident in the evolutionary tracks (Fig.~\ref{fig:zebra_compact}B, red), where upwellings migrate up to 180$^\circ$ from near the substellar point to rise near the antistellar point (180$^\circ$).

A similar degree-1 convection pattern is observed for a planet with a strong lithosphere and a mixed heating mode (Fig.~\ref{fig:evo_compact}D).  The single downwelling is weak since a majority of the lithosphere is sufficiently strong to resist dynamic instability (Fig.~\ref{fig:evo_compact}: D3).  Furthermore, high mantle temperature ($\approx 3488$ K) due to internal heating causes the downwelling to dissipate quickly and also reduces the temperature contrast (hence thermal buoyancy) of upwellings.  Although the downwelling is weak, the high mantle temperature ensures material flows readily due to the temperature dependence of viscosity.  Therefore, a degree-1 flow regime is established with comparable interior velocities as for a purely basally heated planet where the downwelling is stronger but the mantle cooler (compare Fig.~\ref{fig:evo_compact}: B2 and Fig.~\ref{fig:evo_compact}: D2).  The sweeping of hot material to the nightside by the downwelling is evident in Fig.~\ref{fig:evo_compact}(D1) and Fig.~\ref{fig:zebra_compact}(D).

We also investigated if hemispheric tectonics persist for higher nightside temperatures based on the uncertainty from the observations \citep{KKM19}. For basal heating with nightside temperatures of $355$ and $710$ K, we find that hemispheric tectonics (downwellings on dayside) persist (Figs.~\ref{fig:zebra_compact_newmodels}A, B). Similarly, hemispheric tectonics also persist when the internal heating rate is a factor of $2$ less than the reference models (Figs.~\ref{fig:zebra_compact_newmodels}C, D).   

\begin{figure*}
    \centering
    \includegraphics[width=0.85\textwidth]{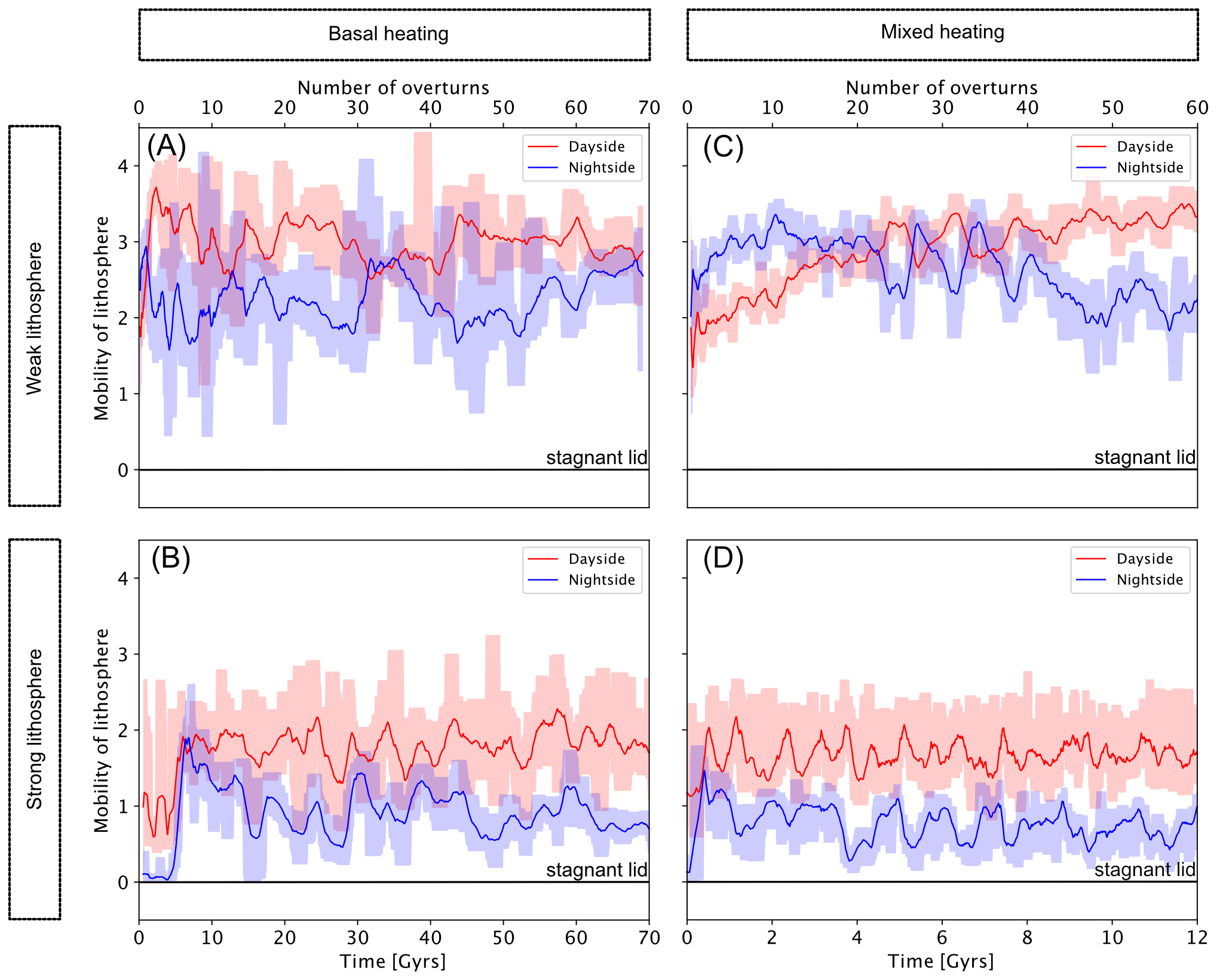}
    \caption{Surface mobility, which is the ratio of the root mean square (rms) surface velocity to the RMS of the convective velocity in the mantle (Earth mobility is around 1.3).  Red and blue lines show the dayside and nightside, respectively.  The shaded regions on either side of the line represent the minimum and maximum values.}
    \label{fig:mobility_compact}
\end{figure*}

\subsection{Hemispheric tectonics: downwellings on the nightside}
We observe another hemispheric tectonic regime if the planet has a weak lithosphere and mixed heating mode (Figs.~\ref{fig:evo_compact}C and \ref{fig:zebra_compact}C).  A degree-1 pattern is established and now strong downwellings occur on the nightside and upwellings accumulate on the dayside (Fig.~\ref{fig:evo_compact}: C3).  Unlike the previous hemispheric regime (a single downwelling on the nightside), there are several downwellings on the nightside as demonstrated by two prominent downwelling tracks either side of the antistellar point (Fig.~\ref{fig:zebra_compact}C, blue) and transient downwellings that occur closer to the day--night terminator.  Weak downwellings form on the dayside, although these tend to remain in the perovskite layer and do not propagate to depth in the post-perovskite layer (Fig.~\ref{fig:evo_compact}: C3, C2).  A prominent upwelling forms at the antistellar point (180$^\circ$) that remains stable and exhibits minimal lateral movement (Fig.~\ref{fig:evo_compact}: C3), anchored by the two prominent downwellings on either side (Fig.~\ref{fig:zebra_compact}C).  The degree-1 pattern of convection is once again driven by downwellings that propagate to depth, in this case on the nightside.  The downwellings sweep hot material in the post-perovskite layer along the CMB toward the dayside, where plumes rise and merge at the substellar point.  A return flow is established in the upper perovskite layer where warm material from the surface of the dayside is advected to the cooler antistellar point (Fig.~\ref{fig:evo_compact}: C2).

\section{Implications} \label{sec:discuss}
\subsection{Hemispheric tectonics and surface mobility}

Our numerical simulations that are constrained by phase curve observations suggest that LHS 3844b may exhibit persistent hemispheric tectonics due to the approximately 1000 K temperature contrast between its day- and nightside.  Hemispheric tectonics are not inferred for any solar system planet at present day.  Currently Earth has a dominant degree-2 pattern of convection, as dictated by downwellings in the Pacific and broad hot anomalies (possibly upwellings) beneath Africa and the Pacific Ocean.  Transient hemispheric tectonics on Earth may occur during supercontinent formation (landmasses assemble on one hemisphere), but according to the geological record this is not a stable and long-lived configuration.  Venus does not have a mobile surface driven by persistent downwellings, but rather its crustal thickness and mean surface age suggests episodic overturn of the surface \citep{RSS18}.  Observations support a prominent role for upwellings to explain surface features (coronae) and active upwellings may cluster in Venus' southern hemisphere \citep{GGM20}.  However, these upwellings are not shepherded into the southern hemisphere by persistent downwellings in the northern hemisphere.

For Mars, a transient degree-1 convection pattern may have initiated following a giant impact that generated massive magmatism to source a Tharsis-like volcanic province \citep{GKG11}.  However, a mobile surface on Mars lasted for only half a billion years or so, after which it transitioned to a stagnant lid (no surface mobility; \citep{ZO16}).  In numerical models of a generic tidally locked Earth-sized planet, \cite{SCG11} observe downwellings on the nightside and upwellings on the dayside; we now find that a similar regime can manifest for a larger rocky planet (LHS 3844b) using phase curve observations and fluid simulations appropriate for a super-Earth interior.  We also discover a previously unrecognized regime of hemispheric tectonics that is characterized by downwellings on the dayside and upwellings on the nightside.  Hence our data-constrained models emphasize the importance of hemispheric tectonic regimes to interpret observations of rocky exoplanets.

For a strong lithosphere, convective stresses in the interior are insufficient to yield the nightside lithosphere and promote downwellings. Instead, the nightside lithosphere establishes a surface return flow to the dayside that accommodates the deep flow of material from the day- to the nightside initiated by downwellings (Figs.~\ref{fig:evo_compact}B, D). As the cold lithosphere is transported to the dayside, it becomes warmer and accommodates more deformation.  All models produce mobile surfaces, so even the strong lithosphere models do not prevent surface motion (Fig.~\ref{fig:mobility_compact}).  Neither the dayside nor nightside transition to a stagnant lid with zero surface velocity (Fig.~\ref{fig:mobility_compact}).  We recover a stagnant lid regime in models (not shown) without plastic yielding (i.e. $\sigma_{\rm duct} \rightarrow \infty$ MPa) and find no interior dichotomy due to the lack of downwellings.  Instead, weak upwellings are approximately uniformly distributed but become stifled by interior heat production that raises the mantle temperature.

We consider lithospheric strength to depend on lithospheric composition, which we assume to be uniform.  Future work could investigate the effects of temperature dependent yield stress.  For our models with a strong lithosphere, the high temperature on the dayside reduces the viscosity of the lithosphere sufficiently to allow it to flow and sink. High temperature would likely decrease the plastic yielding parameter, further facilitating downwellings on the dayside \citep{Hansen2019} and preventing downwellings on the nightside (where the yielding parameter would increase).  Hence the asymmetry between the flow on the dayside and nightside would be further enhanced and increase the likelihood of hemispheric tectonics.  Finally, melting and crustal production would likely increase the overall mobility of the lithosphere \citep{LRT16}.
\subsection{Plumes, volcanism, and observations}
Upwellings (plumes) form through two types of interactions with downwellings \citep[e.g.,][]{TGH02}.  First, the leading edge of downwellings sweep hot material along the CMB to form upwellings (e.g., Fig.~\ref{fig:evo_compact}(B1)).  For hemispheric tectonics, this causes the partitioning of upwellings in one hemisphere and downwellings in the other hemisphere.  Second, upwellings can form beneath downwellings since cold downwellings trap hot material beneath, thus enabling incipient plumes to accumulate buoyancy before breaking through and rising to the surface (e.g., Fig.~\ref{fig:evo_compact}(B3), 30$^\circ$).  A long-lived cluster of upwellings in a particular hemisphere may promote extensive magmatism and volcanism.  For example, upwellings can explain the origin of some large igneous provinces (LIPs) on Earth, which are large surface emplacements of basaltic magma \citep[e.g.,][]{CE94}.  LIPs drive extensive outgassing \citep[e.g. CO$_2$,][]{BG19} and climate modification.  Our models reveal that plumes can be evenly distributed in the interior, preferentially on the dayside, or preferentially on the nightside.  For a uniform distribution, where upwellings are pinned in place by downwellings, upwellings only sample deep mantle material from a small region at the CMB (around 60$^\circ$).  For hemispheric tectonics, plumes are instead flushed around the CMB to the opposite hemisphere, thereby sampling a much larger fraction of the deep interior and delivering this chemical signature to the near-surface.

\begin{figure*}
    \centering
    \includegraphics[width=1.0\textwidth]{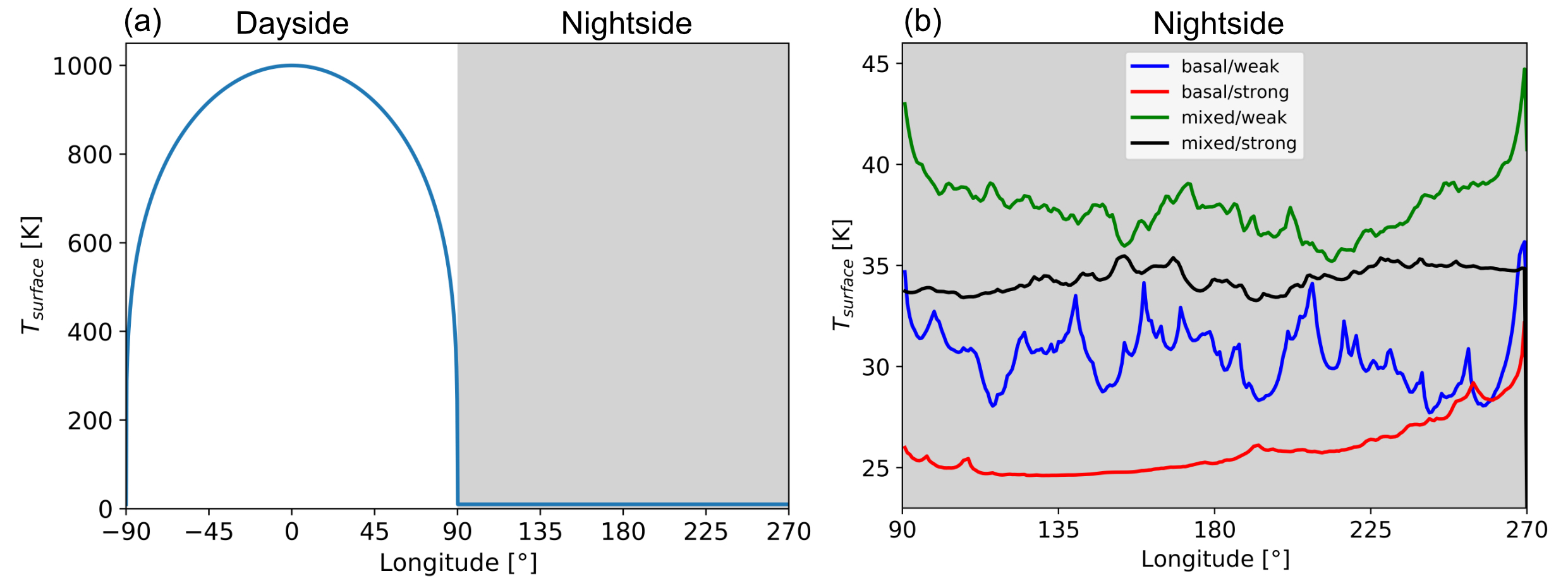}
    \caption{(a) Imposed surface temperature motivated by the thermal phase curve of LHS 3844b. (b) Thermal phase curves of the nightside of LHS 3844b that include the contribution of the heat flux from the interior.}
    \label{fig:phasecurve_total}
\end{figure*}

We time-average the surface heat flux from the interior and determine its contribution to the thermal phase curve.  The thermal phase curve that is applied as the surface boundary condition in our reference models is shown in Fig.~\ref{fig:phasecurve_total}a. Figure~\ref{fig:phasecurve_total}b shows the nightside thermal phase curve including the interior contribution. The dayside phase curve is not shown since the interior heat flux is negligible compared to stellar insolation such that it has little influence on the surface temperature.  The nightside is a favorable target for observing the heat contribution from the interior due to temperatures around $0$ K for an atmosphere-less planet (no reflected light or reradiation of the host star's light). An interior heat contribution, as dictated by the interior convective regime, could produce a variation of the geometric transit depth \citep{Kipping2010}. However, the expected spectral signatures are significantly below the ppm-level in the near-to-mid infrared, which is beyond the capabilities of current or planned instrumentation (Fig.~\ref{fig:phasecurve_total}).

On the dayside, which is observed during the secondary transit, the contribution of the interior heat flux to the eclipse depth would be on the same order of magnitude as for the transit depth. A further complication for the secondary transit is that deconvolving the interior flux from stellar contributions requires knowledge of the stellar spectrum and planetary surface properties (to determine the reflected, absorbed, and reradiated light).  Thus, it may be difficult to observe the contribution from the interior heat to the thermal phase curve if heat is conducted through a solid lid.  However, if massive outpourings of melt occur on the surface associated with plumes (akin to LIPs), this will generate regions of high heat flux and possibly provide a means to test the existence of hemispheric tectonics with future astrophysical observations.  Furthermore, if upwellings facilitate or source outgassing of volatiles on one hemisphere, this may perturb the atmospheric composition and properties on this hemisphere.  This may similarly produce a secondary signal in multiwavelength phase curve observations, in addition to the dominant signal from the day--night temperature contrast.

Parameterized stagnant lid models of volcanic outgassing and atmospheric erosion show that the lack of an atmosphere for LHS 3844b is consistent with a volatile poor mantle \citep{KRU20}.  Our stagnant lid models produce a uniform distribution of upwellings, which implies spatially uniform melting and outgassing at the surface.  Upwellings in the stagnant lid regime are laterally mobile because they are not anchored in place by downwellings.  \citet{KRU20} find that the heat-producing element budget, mantle viscosity, and initial temperature, do not have a significant influence on the size of the atmosphere of LHS 3844b.  However, hemispheric tectonics could lead to differences in melt production and outgassing between the dayside and nightside, potentially modifying the size and chemical composition of an atmosphere.  If LHS 3844b is devoid of an atmosphere and has a volatile poor mantle, we can preclude weakening of the lithosphere by surface water. This suggests LHS 3844b has a strong lithosphere, lending support to hemispheric tectonics characterized by a downwelling on the dayside and upwellings on the nightside.
\section{Conclusions} \label{sec:conc}
Our numerical experiments suggest that hemispheric tectonics may operate on LHS 3844b, where one hemisphere is characterized by downwellings and the opposite hemisphere by upwellings.  For hemispheric tectonics, upwellings may lead to preferential melt generation and outgassing on one hemisphere that could manifest as a secondary signal in phase curve observations. However, the contribution to the thermal phase curve from the interior flux is on the order of 15--30 K, which would produce spectral signatures that are significantly below the ppm level in the near-to-mid infrared, and will therefore be challenging to detect by current and near-future observations. Nevertheless, outpourings of melt (extrusive volcanism) with high temperature ($>1000$ K) fueled by deep mantle upwellings could imprint a signature in the thermal phase curve. If melting is more prevalent on one hemisphere, the associated degassing of volatiles could preferentially perturb the composition and properties of a thin and transient atmosphere on that side of the planet.
\acknowledgments
T.G.M. and D.J.B. acknowledge SNSF Ambizione Grant 173992.  Calculations were performed on UBELIX (\url{http://www.id.unibe.ch/hpc}), the HPC cluster at the University of Bern. T.L. was supported by the Simons Foundation (SCOL award \#611576 to T.L.) and the SNSF (Early Postdoc.Mobility fellowship \#P2EZP2-178621  to T.L.). B.-O. D. acknowledges support from the SNSF (PP00P2-190080).  This research benefited from discussions and interactions within the framework of the National Center for Competence in Research (NCCR) PlanetS supported by the SNSF.  Constructive comments from an anonymous reviewer further sharpened the manuscript.

\bibliographystyle{aasjournal}

\newpage
\begin{thebibliography}{}
\footnotesize
\expandafter\ifx\csname natexlab\endcsname\relax\def\natexlab#1{#1}\fi
\providecommand{\url}[1]{\href{#1}{#1}}
\providecommand{\dodoi}[1]{doi:~\href{http://doi.org/#1}{\nolinkurl{#1}}}
\providecommand{\doeprint}[1]{\href{http://ascl.net/#1}{\nolinkurl{http://ascl.net/#1}}}
\providecommand{\doarXiv}[1]{\href{https://arxiv.org/abs/#1}{\nolinkurl{https://arxiv.org/abs/#1}}}

\bibitem[{Black \& Gibson(2019)}]{BG19}
Black, B.~A., \& Gibson, S.~A. 2019, Elements, 15, 319,
  \dodoi{10.2138/gselements.15.5.319}

\bibitem[{Coffin \& Eldholm(1994)}]{CE94}
Coffin, M.~F., \& Eldholm, O. 1994, Rev. Geophys., 32, 1,
  \dodoi{10.1029/93RG02508}

\bibitem[{Demory {et~al.}(2012)Demory, Gillon, Seager, Benneke, Deming, \&
  Jackson}]{DGS12}
Demory, B.-O., Gillon, M., Seager, S., {et~al.} 2012, Astrophys. J. Lett., 751,
  L28, \dodoi{10.1088/2041-8205/751/2/L28}

\bibitem[{Demory {et~al.}(2016)Demory, Gillon, de~Wit, Madhusudhan, Bolmont,
  Heng, Kataria, Lewis, Hu, Krick, Stamenkovi\'{c}, Benneke, Kane, \&
  Queloz}]{DGW16}
Demory, B.-O., Gillon, M., de~Wit, J., {et~al.} 2016, Nature, 532, 207,
  \dodoi{10.1038/nature17169}

\bibitem[{Golabek {et~al.}(2011)Golabek, Keller, Gerya, Zhu, Tackley, \&
  Connolly}]{GKG11}
Golabek, G.~J., Keller, T., Gerya, T.~V., {et~al.} 2011, Icarus, 215, 346 ,
  \dodoi{10.1016/j.icarus.2011.06.012}

\bibitem[{G\"{u}lcher {et~al.}(2020)G\"{u}lcher, Gerya, Mont\'{e}si, \&
  Munch}]{GGM20}
G\"{u}lcher, A. J.~P., Gerya, T.~V., Mont\'{e}si, L. G.~J., \& Munch, J. 2020,
  Nat. Geosci., 13, 547, \dodoi{10.1038/s41561-020-0606-1}

\bibitem[{Hansen {et~al.}(2019)Hansen, Kumamoto, Thom, Wallis, Durham, Goldsby,
  Breithaupt, Meyers, \& Kohlstedt}]{Hansen2019}
Hansen, L.~N., Kumamoto, K.~M., Thom, C.~A., {et~al.} 2019, Journal of
  Geophysical Research: Solid Earth, 124, 5427,
  \dodoi{https://doi.org/10.1029/2018JB016736}

\bibitem[{Hernlund \& Tackley(2008)}]{HTACK08}
Hernlund, J.~W., \& Tackley, P.~J. 2008, Phys. Earth Planet. Inter., 171, 48,
  \dodoi{10.1016/j.pepi.2008.07.037}

\bibitem[{Kane {et~al.}(2020)Kane, Roettenbacher, Unterborn, Foley, \&
  Hill}]{KRU20}
Kane, S.~R., Roettenbacher, R.~M., Unterborn, C.~T., Foley, B.~J., \& Hill,
  M.~L. 2020, Planet. Sci. J., 1, 36, \dodoi{10.3847/psj/abaab5}

\bibitem[{Kipping \& Tinetti(2010)}]{Kipping2010}
Kipping, D.~M., \& Tinetti, G. 2010, Monthly Notices of the Royal Astronomical
  Society, 407, 2589, \dodoi{10.1111/j.1365-2966.2010.17094.x}

\bibitem[{Kohlstedt {et~al.}(1995)Kohlstedt, Evans, \& Mackwell}]{KEM95}
Kohlstedt, D.~L., Evans, B., \& Mackwell, S.~J. 1995, J. Geophys. Res.-Sol.
  Ea., 100, 17587, \dodoi{10.1029/95JB01460}

\bibitem[{Korenaga(2017)}]{K17}
Korenaga, J. 2017, J. Geophys. Res.-Sol. Ea., 122, 4064,
  \dodoi{10.1002/2016JB013850}

\bibitem[{Kreidberg {et~al.}(2019)Kreidberg, Koll, Morley, Hu, Schaefer,
  Deming, Stevenson, Dittmann, Vanderburg, Berardo, Guo, Stassun, Crossfield,
  Charbonneau, Latham, Loeb, Ricker, Seager, \& Vanderspek}]{KKM19}
Kreidberg, L., Koll, D. D.~B., Morley, C., {et~al.} 2019, Nature, 573, 87,
  \dodoi{10.1038/s41586-019-1497-4}

\bibitem[{Louren\c{c}o {et~al.}(2016)Louren\c{c}o, Rozel, \& Tackley}]{LRT16}
Louren\c{c}o, D.~L., Rozel, A., \& Tackley, P.~J. 2016, Earth Planet. Sci.
  Lett., 439, 18, \dodoi{10.1016/j.epsl.2016.01.024}

\bibitem[{Moresi \& Solomatov(1998)}]{MS98}
Moresi, L., \& Solomatov, V. 1998, Geophys. J. Int., 133, 669,
  \dodoi{10.1046/j.1365-246X.1998.00521.x}

\bibitem[{O'Neill \& Lenardic(2007)}]{OL07}
O'Neill, C., \& Lenardic, A. 2007, Geophys. Res. Lett., 34,
  \dodoi{10.1029/2007GL030598}

\bibitem[{Rolf {et~al.}(2018)Rolf, Steinberger, Sruthi, \& Werner}]{RSS18}
Rolf, T., Steinberger, B., Sruthi, U., \& Werner, S. 2018, Icarus, 313, 107,
  \dodoi{10.1016/j.icarus.2018.05.014}

\bibitem[{Tackley(2000)}]{T00}
Tackley, P. 2000, Geochem. Geophy. Geosys., 1, \dodoi{10.1029/2000GC000036}

\bibitem[{Tackley {et~al.}(2013)Tackley, Ammann, Brodholt, Dobson, \&
  Valencia}]{TAB13}
Tackley, P., Ammann, M., Brodholt, J., Dobson, D., \& Valencia, D. 2013,
  Icarus, 225, 50, \dodoi{10.1016/j.icarus.2013.03.013}

\bibitem[{Tackley(2008)}]{TACK08}
Tackley, P.~J. 2008, Phys. Earth Planet. Inter., 171, 7,
  \dodoi{10.1016/j.pepi.2008.08.005}

\bibitem[{Tan {et~al.}(2002)Tan, Gurnis, \& Han}]{TGH02}
Tan, E., Gurnis, M., \& Han, L. 2002, Geochem. Geophy. Geosys., 3, 1,
  \dodoi{10.1029/2001GC000238}

\bibitem[{Unterborn {et~al.}(2015)Unterborn, Johnson, \& Panero}]{UJP15}
Unterborn, C.~T., Johnson, J.~A., \& Panero, W.~R. 2015, Astrophys. J., 806,
  139, \dodoi{10.1088/0004-637X/806/1/139}

\bibitem[{Valencia {et~al.}(2006)Valencia, O'Connell, \& Sasselov}]{VOS06}
Valencia, D., O'Connell, R.~J., \& Sasselov, D. 2006, Icarus, 181, 545,
  \dodoi{10.1016/j.icarus.2005.11.021}

\bibitem[{van Summeren {et~al.}(2011)van Summeren, Conrad, \& Gaidos}]{SCG11}
van Summeren, J., Conrad, C.~P., \& Gaidos, E. 2011, Astrophys. J. Lett., 736,
  L15, \dodoi{10.1088/2041-8205/736/1/L15}

\bibitem[{{Vanderspek} {et~al.}(2019){Vanderspek}, {Huang}, {Vanderburg},
  {Ricker}, {Latham}, {Seager}, {Winn}, {Jenkins}, {Burt}, {Dittmann},
  {Newton}, {Quinn}, {Shporer}, {Charbonneau}, {Irwin}, {Ment}, {Winters},
  {Collins}, {Evans}, {Gan}, {Hart}, {Jensen}, {Kielkopf}, {Mao}, {Waalkes},
  {Bouchy}, {Marmier}, {Nielsen}, {Ottoni}, {Pepe}, {S{\'e}gransan}, {Udry},
  {Henry}, {Paredes}, {James}, {Hinojosa}, {Silverstein}, {Palle},
  {Berta-Thompson}, {Davies}, {Fausnaugh}, {Glidden}, {Pepper}, {Morgan},
  {Rose}, {Twicken}, {Villase{\~n}or}, \& {the TESS team}}]{VHV18}
{Vanderspek}, R., {Huang}, C.~X., {Vanderburg}, A., {et~al.} 2019, Astrophys.
  J., 871, L24, \dodoi{10.3847/2041-8213/aafb7a}

\bibitem[{Zhang \& O’Neill(2015)}]{ZO16}
Zhang, S., \& O’Neill, C. 2015, Icarus, 265, 187,
  \dodoi{10.1016/j.icarus.2015.10.019}

\end{thebibliography}

\end{document}